\documentclass[fleqn]{POWFI-2018}
\usepackage{wrapfig}
\usepackage{graphicx}
\usepackage{subfigure}

\begin{document}

\ensubject{subject}

\ArticleType{Article}
\SpecialTopic{SPECIAL TOPIC: }
\Year{2019}
\Month{January}
\Vol{60}
\No{1}
\DOI{ }
\ArtNo{000000}
\ReceiveDate{January 11, 2019}
\AcceptDate{April 6, 2019}

\title{Parallel Implementation of $W$-projection Wide-Field Imaging}{Parallel Implementation of $W$-projection Wide-Field Imaging}

\author[1]{Baoqiang LAO}{}%
\author[1]{Tao AN}{antao@shao.ac.cn}
\author[2]{Ang YU}{}
\author[2,3]{Wenhui ZHANG}{}%
\author[2]{\\Junyi WANG}{}
\author[1]{Quan GUO}{}%
\author[1]{Shaoguang GUO}{}%
\author[1]{Xiaocong WU}{}

\AuthorMark{Lao B Q}

\AuthorCitation{Lao et al}

\address[1]{Shanghai Astronomical Observatory, Key Laboratory of Radio Astronomy, Chinese Academy of Sciences, Shanghai, 200030, China}
\address[2]{Guangxi Cooperative Innovation Center of cloud computing and Big Data, Guilin University of Electronic Technology, Guilin, 541004, China}
\address[3]{Guangxi Colleges and Universities Key Laboratory of cloud computing and complex systems, \\Guilin University of Electronic Technology, Guilin, 541004, China}

\abstract{W-projection is a wide-field imaging technique that is widely used in radio synthesis arrays. Processing the wide-field big data generated by the future Square Kilometre Array (SKA) will require significant updates to current methods 
to significantly reduce the time consumed on data processing. {\it Data loading} and {\it gridding} are found to be two major time-consuming tasks in $w$-projection. In this paper, we investigate two parallel methods of accelerating $w$-projection processing on multiple nodes: they are the hybrid Message Passing Interface (MPI) and Open Multi-Processing (OpenMP) method based on multicore Central Processing Units (CPUs) and the hybrid MPI and Compute Unified Device Architecture (CUDA) method based on Graphics Processing Units (GPUs). Both methods are successfully employed and operated in various computational environments, confirming their robustness. The experimental results show that the total runtime of both MPI+OpenMP and MPI+CUDA methods is significantly shorter than that of single-thread processing. MPI+CUDA generally shows
faster performance 
when running on multiple nodes than MPI+OpenMP, especially on large numbers of nodes. The single-precision GPU-based processing yields faster computation than the double-precision processing; while
the single- and double-precision CPU-based processing shows consistent computational performance. The {\it gridding} time remarkably increases when the support size of the convolution kernel is larger than 8 and the image size is larger than 2048 pixels. The present research offers useful guidance for developing SKA imaging pipelines.}



\maketitle

\begin{multicols}{2}

\section{Introduction}\label{sec:1}
The largest international collaboration project in radio astronomy, the Square Kilometre Array (SKA), will be the flagship radio telescope in the next few decades upon completion, providing great opportunities for humans to discover and understand the Universe \cite{SKAsci1,SKAsci2}. The first phase of the SKA (SKA1), consisting of 10\% of the total array, already offers unprecedented performance \cite{SKA1design}. The wide field of view (FoV) , high survey speed, high resolution and high sensitivity of the SKA render vast amounts of observational data. Of particular note is its high survey speed, which is greater than that of the largest existing radio telescope, the Karl G. Jansky Very Large Array (JVLA), by a factor of ~100 \cite{SKA1sci}. The correlated data generated by the low-frequency SKA1 (SKA1-low) will be at a rate of 466 GigaByte per second (GB/s), resulting in 40 PetaByte (PB) of correlated data in a continuous 24-hours of operation. The correlated data will then be imported into the Science Data Processor (SDP) for data processing and calibration in real time. The amount of initially calibrated data from the SDP is dramatically reduced but still remains at the PB level. Deeper analysis ({\it e.g.}, self-calibration and deep imaging) is needed to turn these SDP data into science-ready data products that can be used by scientists. Numerous challenges will be encountered in the processing, curation and storage of the SKA big data \cite{An18}.

Wide-field imaging is one of the main challenges of SKA data processing. Due to the large FoV and non-coplanar baselines of the SKA,
the conventional two-dimensional (2D) Fourier transform approximation in radio synthesis imaging no longer holds, {\it i.e.}, the effect of the $w$-term cannot be neglected \cite{Cor92}. Therefore, new imaging techniques are developed to tackle the wide-field imaging problem of $w$-term correction, e.g., three-dimensional (3D) Fourier transform \cite{SynImg}, faceting \cite{Kog09}, $w$-projection \cite{Cor08}, $w$-stacking \cite{Off14}, and warped snapshots \cite{Cor12} techniques.

Among these techniques, the 3D Fourier transform requires the largest amounts of calculation and memory space, and is thus rarely used in practical applications. The remaining techniques are potentially suitable for SKA imaging, with some being practically used in SKA pathfinder data processing. The faceting algorithm divides the observed sky zone into small pieces named as 'facets'; in each facet the {\it w}-term can be neglected and the 2D Fourier transform can be applied. The produced images in the facets are then combined. There are two types of {\it facets}: in the image-domain \cite{Cor92,Tas18} and uv-domain \cite{Lao18}. One shortcoming of faceting images is that the generated images often suffer from the edge effect induced by misalignment of image brightness scales in adjacent facets. The edge effect and noise interference lead to the reduction of imaging quality.
A recently developed image domain gridding algorithm eliminates the need for computing the convolution kernel and thus reduces the total cost of gridding; moreover it is more accurate compared to classical gridding algorithms \cite{2018A&A...616A..27V}.
The $w$-projection algorithm processes the convolution of the visibility data with a new convolution kernel function (for more details, see Section 2) and then projects the visibility data onto a two dimensional plane, enabling 2D fast Fourier transform (FFT) deconvolution. Although this method also increases the amount of computation and occupies a large amount of computer memory, the created image has higher quality and lower noise level than that obtained by the faceting technique \cite{Cor08}.
The principle of the $w$-stacking technique is to rasterize the visibility data onto different {\it w} layers so as to perform a 2D FFT on each thin {\it w} layer. The processing of $w$-stacking also consumes a large amount of device memory \cite{Pra18}. The warped snapshot technique divides the visibility data according to short time slices; in each slice, visibility data can be assumed to be projected onto a two-dimensional plane.  Each data slice is imaged independently, and the images of all time slices are then combined. The warped snapshot also requires an instantaneously co-planar array and involves multiple FFT calculations. In addition, the warped snapshot technique increases the algorithm complexity and involves additional image stitching time.

Compared with the 2D FFT method, the above techniques can recover the structure of the radio sources located far from the phase center of the image. However, these techniques all involve long computational time and large memory consumption. Therefore, executing these techniques often relies on high-performance clusters or supercomputers. The rapid development of computer technology, especially the development of parallel computing and artificial intelligence technologies, renders parallel execution and acceleration of the wide-field imaging feasible. As the $w$-projection technique produces higher quality images than the faceting technique, has open source code and is easy to transplant \cite{Lao18}, we investigate the single-thread and parallel implementation of wide-field imaging by adopting the $w$-projection technique in this paper.

\section{Parallel Implementation of $W$-projection}\label{experiment}

Parallel optimization of the $w$-projection algorithm has already been presented in literature. Varbanescu et al. introduced a parallel acceleration method based on Cell BE for multicore CPU processors,
with a focus on {\it gridding} and degridding \cite{Var08}.
Although the method accelerated the {\it gridding} and {\it degridding} processing, it increased the amounts of calculations in other steps,
and the overall effect on optimization was not properly evaluated. In addition, the method is implemented on a single node, with an extension to multiple nodes yet to be investigated. Amesfoort et al. \cite{Ame09}
ported the work of Varbanescu et al. to a GPU.
The local optimization problem was partly solved, enabling the bandwidth usage of the device to be maximized, but the limited GPU memory used in their work lowered the imaging resolution, hampering the practical applications to large-scale radio telescope arrays such as the Low Frequency Array (LOFAR) and SKA.
Romein \cite{Rom12} provided a $w$-projection rasterization acceleration method based on GPU architecture, which can reduce the number of memory access, avoid large memory waste, and accelerate the implementation.
Based on Romein's algorithm, D. Muscat developed a GPU-based {\it w}-projection method {\it WImager} which is included in their imaging tool called 'Malta-imager'  \cite{Muscat}.   By utilizing {\it Compression} to merge some consecutive records, {\it WImager} results in 3 times increase of gridding speed; but mention that the {\it Compression} performance depends on observation integration time, grid configuration and the oversampling factor of the convolution functions. They found the GPU-based imaging method is $\sim$100 times faster than a common CPU-based method.
Moreover they found that loading data from disk is a main limiting factor of the execution time.
Simulated data were used in these work, whereas the radio astronomy observational data structure is more complicated, and thus, the practicality of the acceleration algorithm needs further testing.

Though progress has been made in these previous work, more experiments are needed to achieve practical parallelization and acceleration methods of $w$-projection imaging. In particular, parallel implementation and performance test on multiple nodes has not been investigated yet. In this section, we introduce two parallel methods of $w$-projection based on CPU and GPU architectures on multiple nodes.

\subsection{ Parallelizaion Method Using Hybrid MPI and OpenMP Based on CPUs}\label{subsec:3.1}

The {\it data loading} includes the time consumed on (1) reading visibility data and auxiliary data contained in Measurement Set files from external storage media to disk on clusters or HPCs, (2) copying the visibility data from disk to memory, and then (3) copying data from memory to CPU or GPU. In the CPU-based pipeline, the {\it data loading} only contains steps (1) and (2). These processes of transferring data are found to be highly dependent on the I/O bandwidth of the storage device (e.g., \cite{Muscat}) and cost a large amount of execution time since the {\it gridding} process can only commence after all data is loaded into memory. The parallelization of GPU CUDA program requires that the visibility data are time ordered. Thus, additional re-ordering time is spent on the {\it data loading} in the GPU-based processing pipeline, compared with CPU-based pipeline.
There are basically two approaches to perform parallel implementation to accelerate {\it data loading}. One is to parallelize the data I/O in the underlying layer, and the other is to write a parallel {\it data loading} program in the application layer before data processing.

The Casacore Table Data System (CTDS) \cite{CASAcore} is the data I/O library of the Measurement Set. Currently, CTDS only allows data reading from and writing to a table of the Measurement Set in serial mode. This will eventually consume a large amount of time when the data size of the Measurement Set becomes very large. CTDS provides an interface for users to design and implement their own CTDS storage managers. Wang et al. developed the Adaptive I/O System (ADIOS) based on the parallel storage management called AdiosStMan \footnote{https://github.com/SKA-ScienceDataProcessor/AdiosStMan}, which is the first parallel I/O system for the Measurement Set \cite{Wang16}. The performance test shows that AdiosStMan has an excellent scalability on up to 80 compute nodes for parallel writing. AdiosStMan is currently in the experimental stage and has not been widely used in practical astronomical data processing. Therefore, we choose the second parallel approach for {\it data loading}. In our method, we use MPI to implement parallel {\it data loading} and allocate parallel data processing on multiple processors and multiple compute nodes.

In {\it gridding} step, the original $w$-projection imaging provides a coarse parallelization approach. As shown in Fig. 1, the {\it gridding} step is embedded in a loop process, thus providing a natural way to accelerate the processing by parallelizing that loop operation. The parallelization can be achieved using OpenMP, which is a parallel standard form of shared memory that can achieve parallel optimization without complex code porting \cite{Dag98}. More details are presented in \textcolor{red}{Appendix B}.

\begin{figure}[H]
\centering
\includegraphics[scale=0.6]{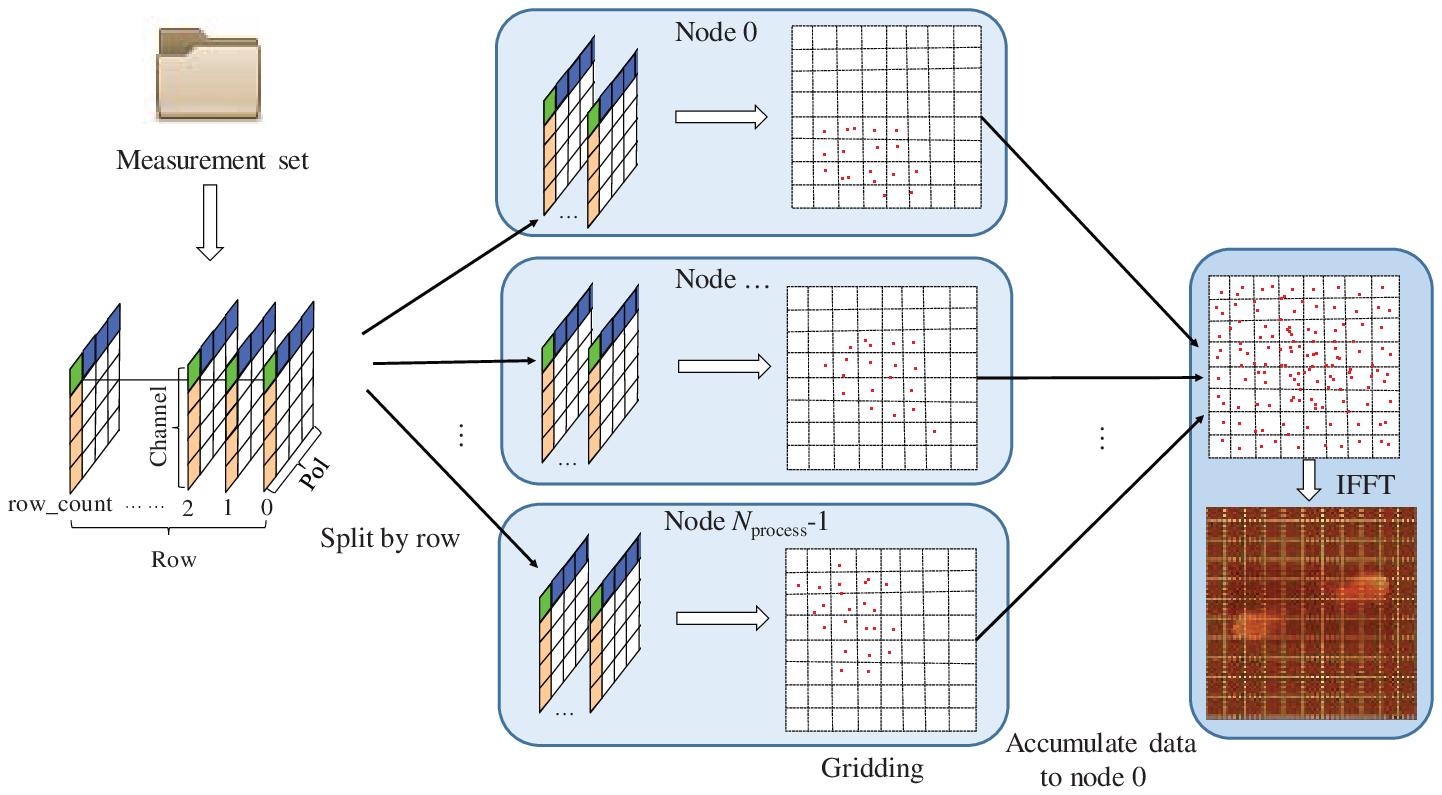}
\caption{
Sketch map of the hybrid MPI+OpenMP parallel procedure.The visibility data read from the target field of a Measurement Set are stored into a 3-dimensional array, including row, frequency channel and polarization ({\it pol}). The  'row\_count' in this figure is $N_{\rm obs\_time\_range} \times (N_{\rm baselines}-1) $). The visibility data will be first split by row using MPI processes. Next each data slice is gridded in certain sub-grids on the full uv plane using multi-threads by OpenMp. Then all gridded data are accumulated onto the whole uv plane and sent to the master node  ({\it e.g.} node 0). Finally, an IFFT is performed with the gridded data to create an image on node 0.
}
\label{fig:fig1}
\end{figure}

\cref{fig:fig1} shows the flow chart of the hybrid MPI+OpenMP parallel strategy running on multiple compute nodes, which can be multiple CPU nodes or GPU nodes with multiple CPU cores. First, the input Measurement Set data are split into pieces using MPI, where $N_{\rm process}$ is the number of MPI processes. Data processing of each piece of data is executed on a number of compute nodes, each of which is identified by the MPI rank number. Second, OpenMP allocates multiple threads to parallelize {\it gridding} loops of each piece of data on the allocated compute node. Each thread is assigned $N_{\rm iteration}/N_{\rm thread}$ iterations. Third, the accomplished {\it gridding} results on each compute node are accumulated and sent to the master node (i.e., node \#0). Finally, the master node performs an IFFT on the transmitted {\it gridding} result to obtain the final image.

\subsection{ Parallelization Method Using Hybrid MPI and CUDA Based on GPUs}\label{subsec:3.2}

\begin{figure}[H]
\centering
\includegraphics[scale=0.6]{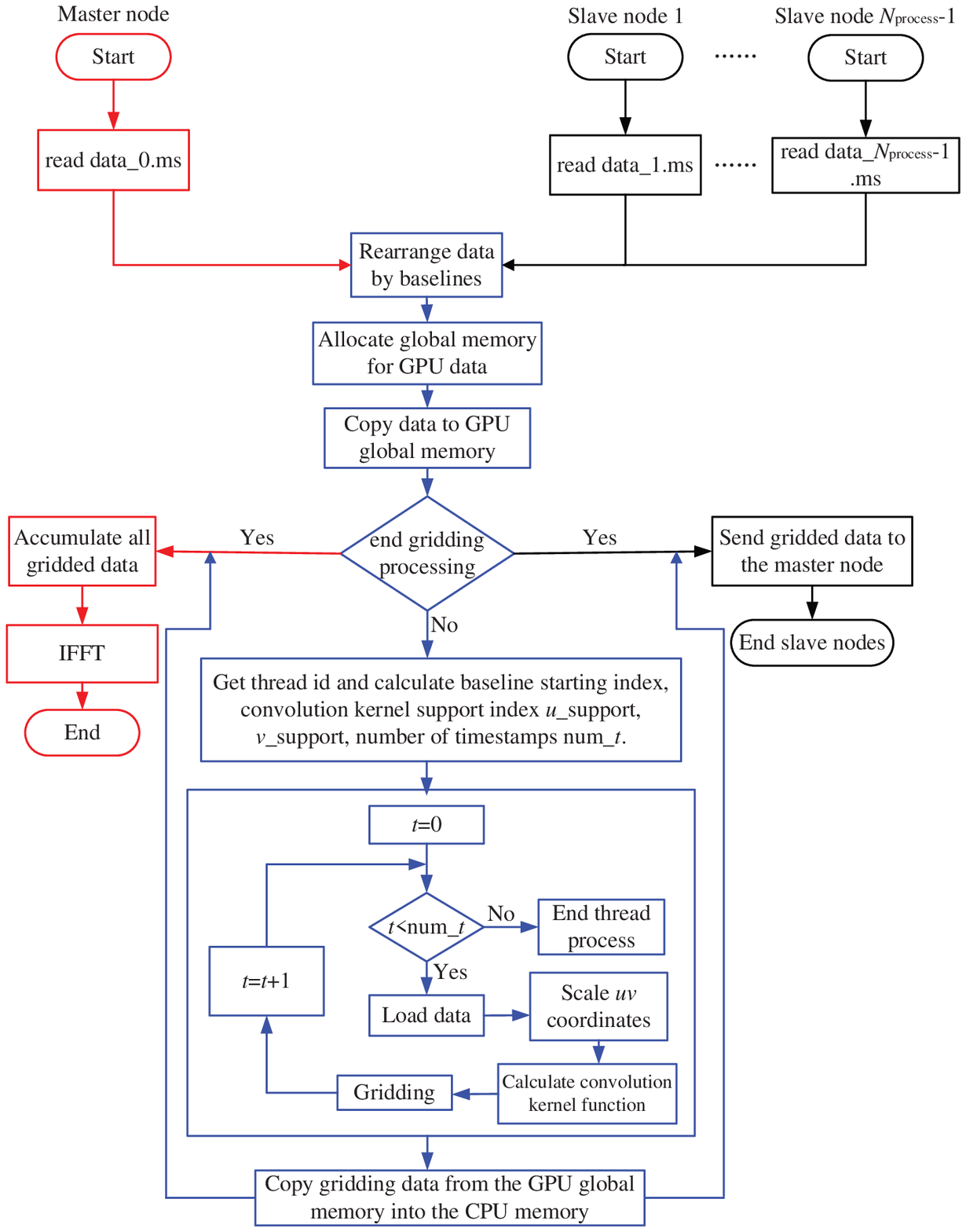}
\caption{Flow chart of the hybrid MPI+CUDA parallelization procedure based on GPUs. The total number of CUDA threads is expressed as ${N_{{\text{Baselines}}}} \times {C_{{\text{Full}}\_{\text{support}}\_{\text{size}}}} \times {C_{{\text{Full}}\_{\text{support}}\_{\text{size}}}}$. The size of the thread block is 256. The start index of the current baseline and the support index of the convolution kernel are defined as $u$\_support and $v$\_support, respectively, and the number of timestamps (num\_$t$) on each baseline is calculated. The convolution {\it gridding} is performed according to the timestamp loop, and the gridded data of the current baselines are cached into the CUDA register. Each thread only computes the accumulation result under the support index of the current convolution kernel. The current baseline gridded data will be copied from the CUDA register to the GPU global memory when the data processing moves to the next baseline. Finally, the gridded data temporarily saved in the GPU global memory are copied into the CPU memory when all {\it gridding} processing threads are completed, and then an IFFT is used to create the final image.}
\label{fig:fig2}
\end{figure}

In {\it data loading} on GPU architecture, we also use MPI to parallelize {\it data loading} and processing but to split data by time. Figure \ref{fig:fig2} shows the parallel procedure based on hybrid programming of MPI and CUDA. The Measurement Set file should be split into multiple sub-Measurement Set files by observation time using the CASA SPLIT task in advance. The number of sub-Measurement Set files is equal to the number of MPI processes.
More details of the {\it data loading} and splitting are presented in \textcolor{red}{Appendix B}.

We should note that the premise of parallelization using the CUDA program is that the observational data have been grouped by baseline and then ordered by time. However, the observation data are
not
stored in this order. Therefore, extra data repackaging and sorting is necessary in the {\it data loading} step before data processing starts. In addition, the
number
of observation time range (time periods) on all baselines are not uniform. This may happen when data in some time range on some certain baselines are deleted from the Measurement Set file, being identified as radio frequency interference (RFI) singals or other observation and/or calibration errors. Therefore, a variable-length array is used to store the data on each baseline, but this variable-length array is very difficult to transfer to the GPU for calculation. We thus use a one-dimensional array to store the first scan position (called "timestamp" in the imaging pipeline) of each baseline. The Measurement Set does not store the baseline index; only the ID number of each antenna is stored. We provide a method of calculating the baseline index as follows. First, the data are grouped by baseline, and each group corresponds to a number, which can be calculated using the ID number of the antenna:
\begin{equation}
{I_{\bm B}} = \left( { - {S^2} + 2S{N_{{\text{ant}}}} + S} \right)/2 + \left| {{I_{{\text{ant1}}}} - {I_{{\text{ant2}}}}} \right|{\mkern 1mu}
\label{eq:5}
\end{equation}
where ${\bm B}$ represents the baseline consisting of antenna 1 (ant1) and antenna 2 (ant2), $I_{{\text{ant1}}}$ and $I_{{\text{ant2}}}$ are the ID numbers of ${{\text {ant1}}}$ and ${\text {ant2}}$, and $S$ is the minimum ID number of ${\text {ant1}}$ and ${\text {ant2}}$.

Then, the baseline number is calculated according to \cref{eq:6} and count the number of timestamps under the same baseline number ({\it i.e.}, the number of samples of the same baseline) and store them in the array $C$. Finally, use the prefix sum method to calculate the starting index of each baseline; its mathematical expression is
\begin{equation}
{\bm B_{start\_index}}[{I_{\bm B}}] = \left\{ \begin{gathered}
  0\quad \quad \quad \quad \;\;{I_{\bm B}} = 0 \hfill \\
  \sum\limits_{k = 0}^{{I_{\bm B}} - 1} {C[k]} \quad \quad {I_{\bm B}} > 0 \hfill \\
\end{gathered}  \right.
\label{eq:6}
\end{equation}
Combining the start value of the baseline index calculated above and the number of timestamps under the same baseline index, the imported data can be reordered according to the baseline index.

A GPU, having thousands of compute cores, has great advantage compared to a CPU in parallel computing. Therefore, it is more efficient to accelerate the {\it gridding} step with a GPU. Here, we introduce the implementation method of {\it gridding} acceleration by using CUDA. Because the integration time between consecutive measurements is short enough, continuous measurements on the same baseline will be gridded onto the same grid point. The {\it gridding} results of multiple timestamp data continuously measured on the same baseline are then accumulated and buffered in a register. This feature allows {\it gridding} processing to be effectively combined with CUDA's threading model. The data in the register are written into the global memory when the processing moves to the next baseline data. There is no competition between multiple threads in this way. Parallel {\it gridding} on multiple baselines can be performed if the write access to the global memory is complete in an {\bf atomic} manner. In the {\it gridding} process, the operation of each convolution kernel is assigned as a thread, and the entire convolution process requires a total of ${C_{{\text{Full}}\_{\text{support}}\_{\text{size}}}} \times {C_{{\text{Full}}\_{\text{support}}\_{\text{size}}}}$ threads, where ${C_{{\text{Full}}\_{\text{support}}\_{\text{size}}}}$ is the full support size of the convolution kernel. The minimum number of threads for the CUDA programming is ${N_{{\text{Baselines}}}} \times {C_{{\text{Full}}\_{\text{support}}\_{\text{size}}}} \times {C_{{\text{Full}}\_{\text{support}}\_{\text{size}}}}$, where ${N_{{\text{Baselines}}}}$ is the total number of baselines.

In \cref{fig:fig2}, the processes bounded by the red boxes indicate the processes running on the master node, those bounded by black boxes indicate the processes running on the slave nodes, and those bounded by blue boxes indicate the processes running on the both master and slave nodes.

The visibility data first are imported from sub-Measurement Set files stored on the master compute node and slave compute nodes in parallel. Then, the visibility data will be re-ordered and re-packaged (using the method discussed above) on each compute node. After re-ordering, the data are copied to GPU memory which is allocated to store those data before copying. Next, {\it gridding} will be executed on each compute node in parallel manner by using the CUDA program. The exported gridded data from each slave compute node will be sent to the master node, where all gridded data are combined into a whole data set. A final fast IFFT operation is performed to create the image.

\section{Results and Discussion}\label{result}

\begin{figure*}[t]
\includegraphics[width=0.58\textwidth]{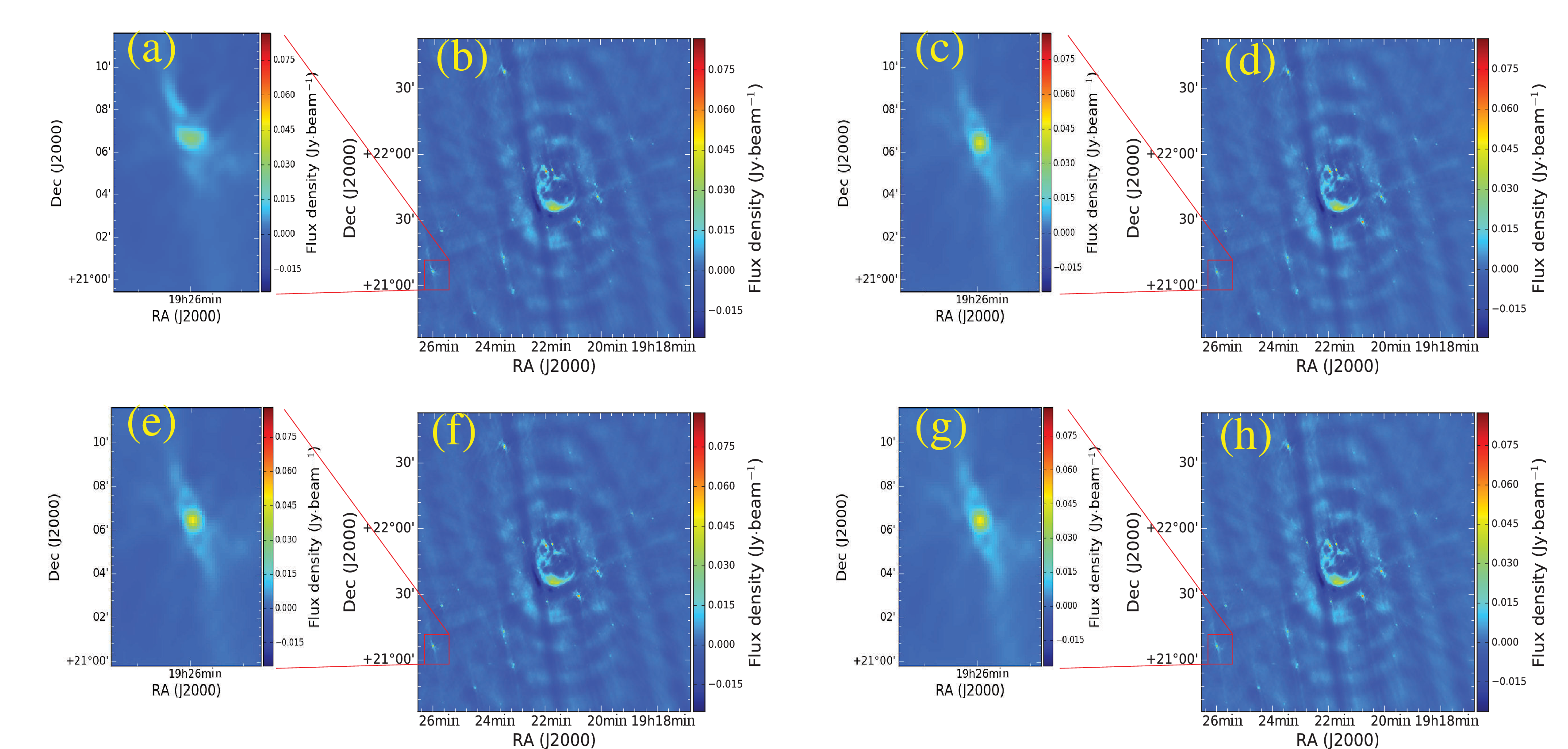}
\label{fig:fig3}
\caption{Images obtained from four methods. (b) traditional two-dimensional fast Fourier transform method; (d) $w$-projection processing using a single thread; (f) MPI+OpenMP method based on CPU architecture; (h) MPI+CUDA method based on GPU architecture. (a), (c), (e) and (g) are the test off-center point source of (b), (d), (f) and (h) near RA=19 h 26 m, DEC=$21^{\rm o}7^{'}$, respectively.}
\end{figure*}

In this section, we will present the results from three test experiments: validation, robustness and scalability. The implementation, computer configuration and results will be presented in detail for each experiment.

\subsection{Validation Test}\label{subsec:3.1}
The purpose of this experiment is to validate the correctness of our methods as well as the programs by using observational data. This experiment was conducted on the Tianhe-2 (MilkyWay-2) supercomputer \cite{Liao14}.

Tianhe-2 was the world's fastest supercomputer in the TOP500 list \footnote{http://www.top500.org/} from 2013 until 2016 and is now the fourth fastest supercomputer. Tianhe-2 has a total of 17920 compute nodes, among which 25 are GPU nodes and the remainder are CPU nodes. \textcolor{red}{Appendix Table B1} lists the specification of a single GPU node on Tianhe-2. Each node has two CPU chips (10 cores each) and two GPU cards (NVIDIA Tesla K80).

The observational data used in this experiment were observed with the JVLA \footnote{https://casaguides.nrao.edu/index.php/EVLA\_Wide-Band\_Wide-Field\_Imaging:\_G55.7\_3.4-CASA4.4\label{web:3}}. Those observations were made at L band (covering the 1-2 GHz frequency range) in the D-array configuration from 01:00:25 to 08:25:16 on August 23, 2010, with a duration of approximately 8 hours. The data size is 15 GB. The purpose of these JVLA observations is to validate the wide-field imaging methods integrated in the CASA software. Moreover, the JVLA is an interferometer similar to SKA1-mid. Therefore, these data are suitable for verifying the SKA1-mid data processing techniques. We first run calibration tasks on the data using CASA software following the procedure given in \ref{web:3}{\color{blue}{)}} to prepare the calibrated data for the input data of our tests.

The validation experiment used 10 GPU nodes of Tianhe-2. In the test of the hybrid MPI+OpenMp method, we used the 10 GPU nodes but did not enable the GPU card. Each node contains 20 CPU cores, so actually we used 200 CPU cores in total. In the test of the hybrid MPI+CUDA method, one GPU card on each of the ten GPU nodes was enabled, so a total of 10 GPU cards were used. The primary argument settings for these tests are as follows: the final output image is 1024 pixel $\times$ 1024 pixel, the cell size of the image is 8 arcsecond in the right ascension (RA) and declination (DEC) directions.
The support size of the convolution kernel depends on the minimum angular scale to be sampled in the image plane \cite{Tas13}. Higher spatial resolution imaging requires larger support size of the convolution kernel, though with a vast computation cost. In addition, for wide field of view and non-coplanar array, large support size of convolution kernels is needed to describe the image features. A reasonable selection of the support size is a compromise between computation cost and observation configuration.
In our experiments, the number of $w$-projection planes is 23, and the full support size of the convolution kernel is 9.

The results are shown in the bottom panel of Figure~\ref{fig:fig3}. As a comparison, the results from a 2D FFT and single-thread $w$-projection are demonstrated in the top panel.2D FFT The shell-like structure of the supernova remnant G55.7+3.4, which is prominent in the image centers, does not show significant difference in the four panels. However, the inset of the 2D FFT image (top left) clearly shows a distorted morphology of a point source far away from the image center, near RA=19 h 26 m, DEC=$21^{\rm o}7^{'}$, when compared with other panels. Other point sources close to the image edge also suffer from distortions at various levels. The images produced by MPI+OpenMP (bottom left) and MPI+CUDA (Figure~\ref{fig:fig3}-f and h) are quantitatively
consistent with each other and also with that acquired from the single-thread CASA $w$-projection (Figure~\ref{fig:fig3}-d).   The noise level of each image is 2.2, 2.3 and 2.3 mJy beam$^{-1}$, respectively; the peak brightness of a test off-center point source shown in the inset panels Figures \ref{fig:fig3}-c, e and g is 45, 48, and 49 mJy beam$^{-1}$, respectively,  and  they are consistent within 2$\sigma$.

To assess the acceleration of the parallelization methods, we ran each test ten times and recorded each runtime. The average runtime was calculated. The runtime of the single-thread processing is 492.84 seconds. The runtime of the hybrid MPI+OpenMP is 3.59 seconds on a total 200 CPU cores or with 200 threads, and the hybrid MPI+CUDA takes \textbf{2.73} seconds on 10 GPU nodes (one GPU card each node). It is obvious that MPI+OpenMP and MPI+CUDA are faster than the single thread by factors of \textbf{137.3} and \textbf{180.5}, respectively, confirming that the parallel processing indeed accelerates the $w$-projection processing. MPI+CUDA is \textbf{1.3} times faster than MPI+OpenMP method.
The GPU acceleration performance is consistent with the previous experiments, e.g.,  \cite{Muscat}.
More comparison of the operation results on CPU- and GPU-architecture will be presented in Section 4.3.

\subsection{Performance Test of the Parallelization Methods}\label{subsec:3.2}
The motivation of the performance test is to demonstrate the robustness of parallel acceleration programs on various compute environments and configurations and to explore the scalability from a single node to multiple nodes.

\textbf{Robustness tests on single node}

The experiments were performed on three clusters with diverse configurations: Tianhe-2 supercomputer (Tianhe-2, in short), the SKA China Data Center prototype at the Shanghai Astronomical Observatory (SKACDC, \textcolor{red}{see Appendix Table B2}), and the Galaxy supercomputer of the Australian Pawsey Supercomputing Center (Pawsey, \textcolor{red}{see Appendix Table B3}).

\begin{figure}[H]
\centering
\includegraphics[scale=0.43]{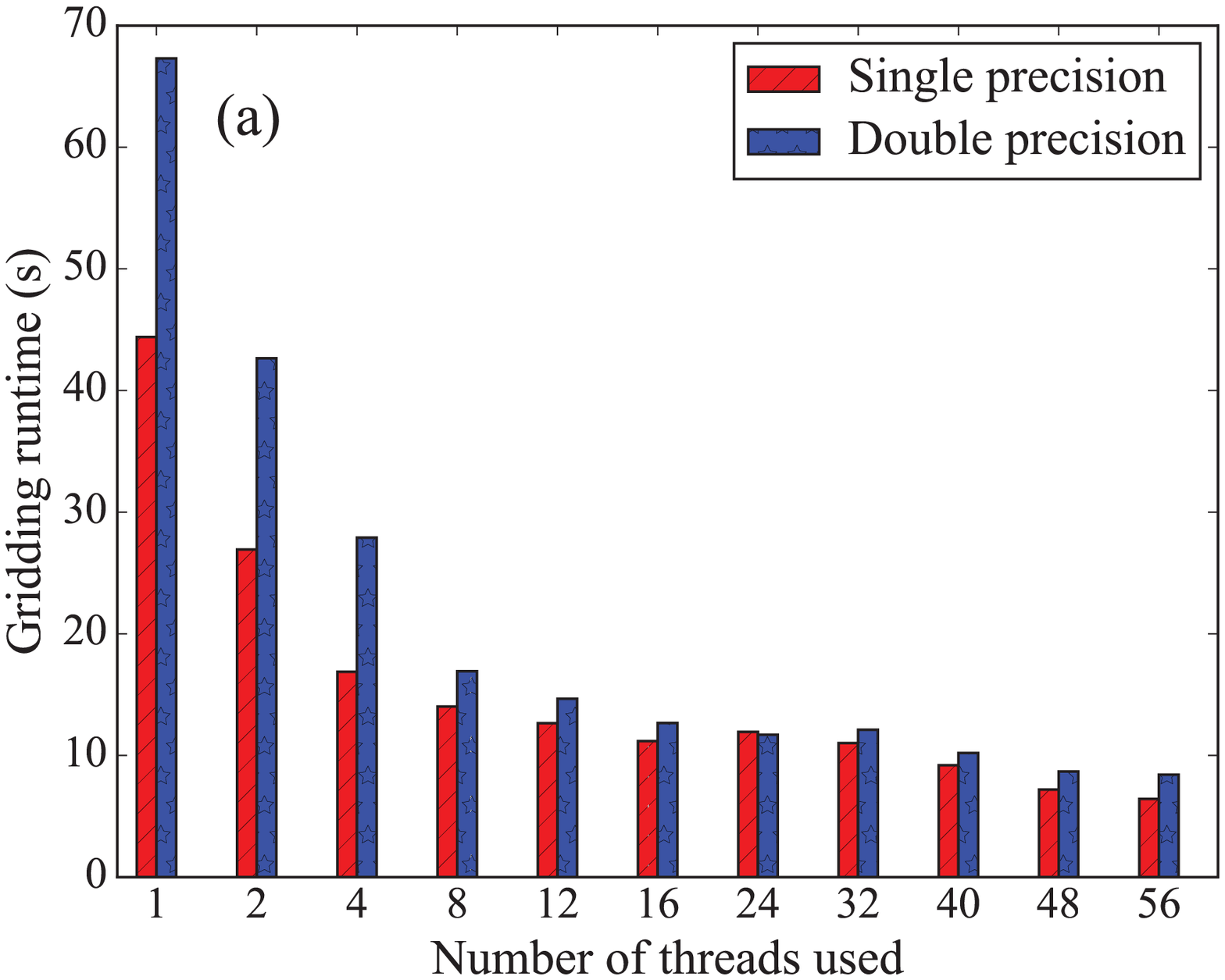}
\vskip15pt
\includegraphics[scale=0.43]{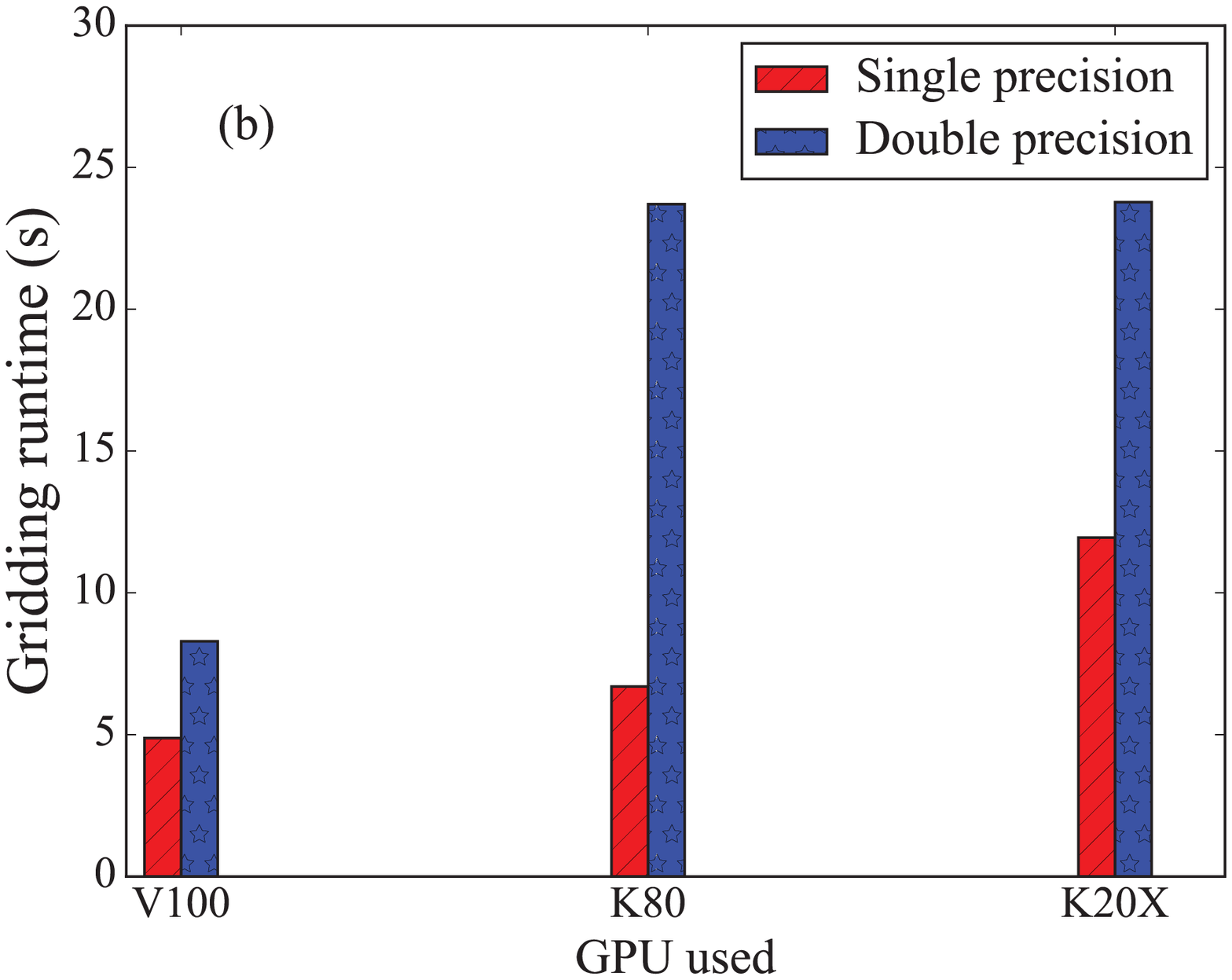}%
\caption{The results of performance tests on a single node. Top panel: {\it gridding} running on China SKA Data Center prototype cluster using upto 56 CPU cores,  Bottom panel: {\it gridding} operation on three different-type GPU cards.}
\label{fig:fig4}
\end{figure}

One GPU node on each platform was used for the tests. The GPU cards are of different types. \cref{fig:fig4} compares the {\it gridding} runtimes on diverse GPU machines.
The Tesla V100 runtimes show the best computational performance in either single or double precision case. The K80 GPU is 2 times faster than the K20X GPU in the single precision mode, but their runtimes in the double precision mode are similar. This experiment confirms that the parallel $w$-projection program can be deployed on diverse cluster environments.

Of note, the experiment undertaken on SKACDC is among the performance tests of the prototype machine. In this test, we also used CPUs of the GPU node of the SKACDC cluster to investigate the performance of the algorithm of MPI+OpenMP on a single node; each of the two CPUs has 56 cores. The number of threads is controlled by changing the environment variable OMP\_NUM\_THREADS. The results are compared in \cref{fig:fig4}. It displays the runtimes of the most time-consuming task, {\it gridding}. The white and black bars represent the single and double precision experimental results, respectively. In all cases, the single precision costs less time than double precision. The same result is seen in other experiments in this work as well. The {\it gridding} runtime decreases as the number of threads (CPU cores) increases in both single and double precision modes, but the scaling is not linear. The {\it gridding} runtime when using a single thread is 6.9 times that when using the maximum 56 threads. The curves flatten after 8 threads (CPU cores).
The change is most likely due to an imbalance in the workload that the number
of rows of visibility being gridded cannot be divided evenly between threads of the MPI+OpenMP method. This remains an open question to be solved in future work. The runtime for the V100 GPU is shorter than for 56-core CPU processing, suggesting that {\it gridding} using a GPU is better than that using a multicore CPU, in particular when CPU threads used are small.

The above results demonstrate that both MPI+CUDA and MPI+OpenMP work normally on a single node under various experimental environments.

\textbf{Scalability tests on multiple nodes}

\begin{figure}[H]
\centering
\includegraphics[scale=0.6]{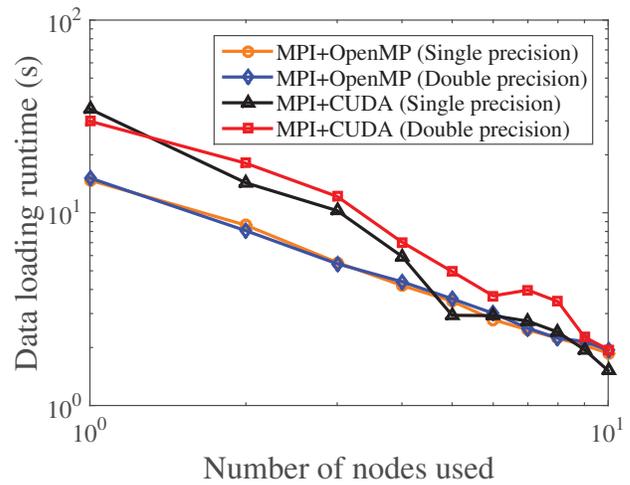}
\caption{Change of {\it data loading} time with number of nodes.
This experiment was made using 10 compute nodes of Tianhe-2 Supercomputer.
Runtimes of the four sets of processing results are shown with different colored symbols.}
\label{fig:fig5}
\end{figure}

\begin{figure}[H]
\centering
\includegraphics[scale=0.6]{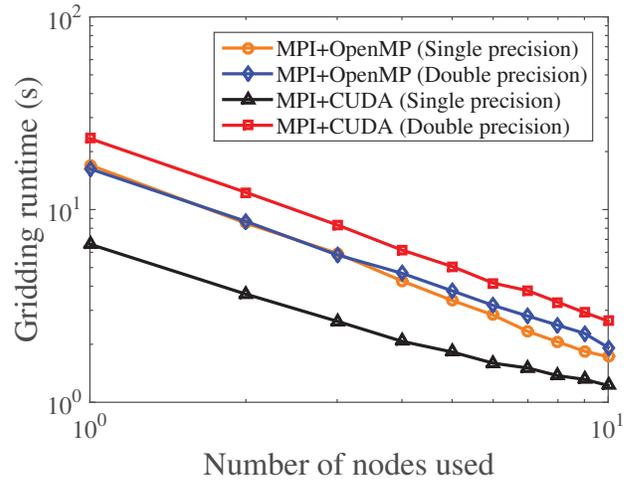}
\caption{Change of {\it gridding} time with number of nodes.}
\label{fig:fig6}
\end{figure}

We then carried out multiple-node tests with the purpose of investigating how performance changes with number of nodes. This experiment was deployed on Tianhe-2 using the same setup as the experiment in \cref{subsec:3.1}. Ten tests were conducted, with the number of nodes increasing from 1 to 10 in sequence. \cref{fig:fig5} and \cref{fig:fig6} present the runtimes recorded by {\it data loading} and {\it gridding}, respectively.

In \cref{fig:fig5}, the {\it data loading} times for both the MPI+OpenMP and MPI+CUDA methods show consistent decreasing trends with increasing numbers of compute nodes. The single and double precision processing curves do not exhibit significant difference.
The runtime decreases rapidly by a factor of 3.5-5.8 as the number of compute nodes increases from $N_{\rm {node}}$=1 to $N_{\rm {node}}$=4 (\textcolor{red}{see Appendix Table B5}), showing a roughly linear relationship between runtime and number of compute nodes. However, the runtime - number of nodes curve flattens after $N_{\rm {node}}=5$.

We found a notable difference between MPI+OpenMP and MPI+CUDA curves at small $N_{\rm {node}}$; the {\it data loading} runtime using GPU processing is much longer than that for CPU processing. Quantitatively, the {\it data loading} time for single-precision MPI+OpenMP processing is approximately 14.8 seconds, but the single-precision MPI+CUDA takes 34.6 seconds. The longer {\it data loading} time in GPU-based processing actually arises from the extra time consumed on
copying data from disks of host machine to GPU global memory
(see Section 3.2 and Figure 4).
Muscat \cite{Muscat} also found that loading data from disk is the main limiting factor of their GPU acceleration experiments.
The MPI+CUDA runtime becomes comparable to or slightly shorter than that of MPI+OpenMP until $N_{\rm node}$=8. Further tests on even more nodes are necessary to explore whether the GPU-based processing runtime is smaller than the CPU-based runtime on larger numbers of nodes.

\cref{fig:fig6} shows the runtime - number of nodes curves of the {\it gridding} task derived from CPU-based (MPI+OpenMP) and GPU-based (MPI+CUDA) processing. As with the curves in \cref{fig:fig5}, the runtime gradually decreases as the number of employed nodes increases, verifying the scalability of the MPI+OpenMP and MPI+CUDA methods in the parallel {\it gridding} task. The slopes are steeper at small $N_{\rm node}$ and then become flatter at large $N_{\rm node}$. The varying slopes can be interpreted employing the same reasoning used to interpret the {\it data loading} time curves above. Floating-point precision exerts little influence on runtime in the CPU-based method, MPI+OpenMP processing. However, the GPU-based MPI+CUDA shows large discrepancies for all node cases, and the single-precision runtime is approximately 1/4 of the double-precision time for $N_{\rm node}<=3$, changing to approximately 1/2 for $N_{\rm node}>=9$. This suggests that single-precision floating-point computation is faster than double-precision computation on GPU architecture. This is a useful implication because radio astronomical data are often kept with single precision.

In summary, the scalability of the MPI+OpenMP and MPI+CUDA methods is verified. The GPU-based MPI+CUDA processing, in general, displays a superior advantage of parallel acceleration using large supercomputers with multiple nodes. The single-precision MPI+CUDA computation is fastest in the {\it gridding} operation.

\vspace{2ex}
\textbf{Impact of parameter selection on parallel implementation}

We further analyze the impacts of two key parameters, the support size of the convolution kernel and the image size, on the parallel implementation. In these tests, we used 10 GPU nodes of the Tianhe-2 supercomputer and took {\it gridding} as an example.

\begin{figure}[H]
\centering
\includegraphics[scale=0.6]{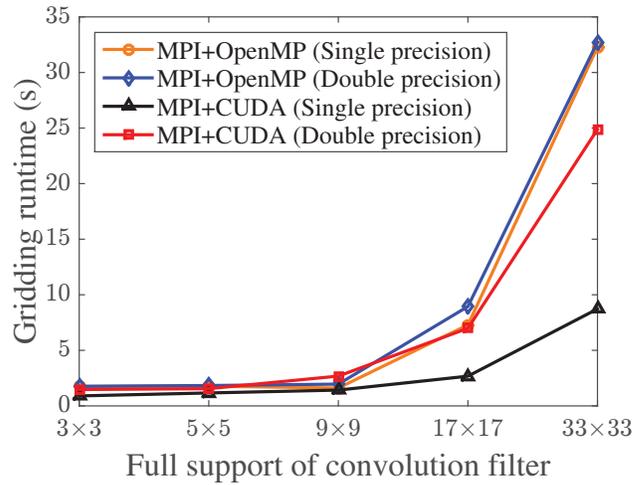}
\caption{Change of {\it gridding} time with the full support size of convolution kernel.} \label{fig:fig7}
\end{figure}

\cref{fig:fig7} shows the changes in {\it gridding} time with the support size of the convolution kernel. As the full support size increases from 3 to 9, the {\it gridding} time steadily increases. We can see from \cref{fig:fig2} that large support sizes of the convolution kernel result in more loops in the blue-colored boxes, thus increasing the {\it gridding} runtime. A sharp increase in {\it gridding} time is seen when the support size is higher than 9, implying that a proper selection of convolution kernel size is important for saving the total runtime. MPI-CUDA processing appears to be faster than the MPI+OpenMP method. This advantage becomes even prominent with a large number of convolution kernels.

\begin{figure}[H]
\centering
\includegraphics[scale=0.6]{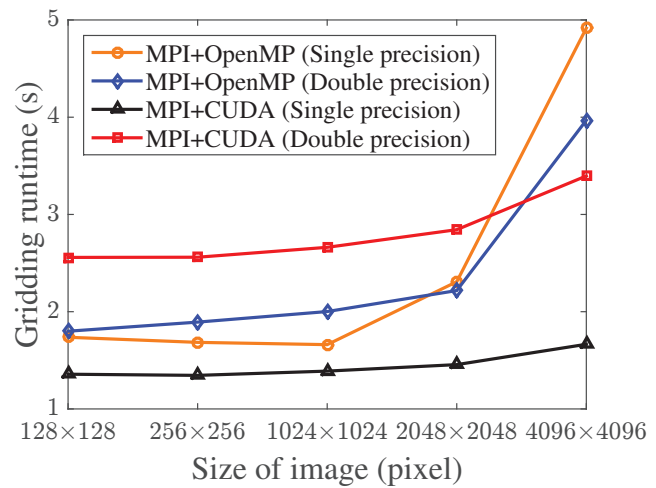}
\caption{Change of {\it gridding} time with the image size.}
\label{fig:fig8}
\end{figure}

\cref{fig:fig8} shows the variation of {\it gridding} time with image size. As seen from the figure, the {\it gridding} time steadily increases as the image size becomes larger until it reaches 2048 pixel $\times$ 2048 pixel. A sharp jump occurs from 2048 to 4096 pixels, and the robustness becomes poor. Measurements of this sharp transition should be further investigated to extend the scalability to high-resolution and large-size images. The percentage increment in the MPI+OpenMP processing is higher than that in the MPI+CUDA processing, suggesting that the scalability of the GPU-based algorithm is more robust. Single-precision processing in either the CPU-based or GPU-based method costs less time than the corresponding double-precision processing, in agreement with the results derived from the experiments above.

\section{Summary and Conclusions}\label{summary}
In this paper, we investigated two methods for parallel acceleration of the $w$-projection wide-field imaging technique. One is a hybrid MPI and OpenMP method based on CPU architecture, and the other is a hybrid MPI and CUDA method based on GPU architecture. The main conclusions of the research are summarized as below.

(1) Applying $w$-projection effectively recovers the image distortion at the edge of the large field of view. The image quality derived from the MPI+OpenMP and MPI+CUDA is quantitatively consistent with that derived from single thread processing, suggesting that the parallelization is successful and does not include any artifacts. Moreover, the parallel processing is 137-181 times faster than the single thread processing.

(2) Two methods are successfully tested on various compute environments on the Tianhe-2 supercomputer (China), the Pawsey supercomputer center (Australia) and the SKA China Data Center (SKACDC) prototype, verifying the robustness of the programs. The SKACDC also provides a demonstration of its advanced performance, enabling radio interferometer data processing.

(3) Two time-consuming tasks, {\it data loading} and {\it gridding}, are chosen to test the parallel scalability on multiple nodes. In general, {\it data loading} and {\it gridding} runtime decrease with the number of nodes in MPI+OpenMP and MPI+CUDA processing, confirming the scalability of the methods. It also suggests that the parallel programs are adaptive to large-scale deployment on clusters or supercomputers. We should mention that the "runtime - number of nodes" curves show linear relationships only at small number of nodes. The slopes become flatter with an increasing number of nodes. The flattening of the curves results from the imbalance in the workload between the multiple threads and nodes. This would be a problem that should be seriously considered in large SKA imaging operation, especially when there are multiple concurrent tasks of diverse science applications.
Future experiments by integrating the {\it image domain gridding} in the pipeline  would be desirable to further reduce the runtime of {\it w}-projection.

(4) MPI+CUDA spends more time on {\it data loading} than MPI+OpenMP on a small number of nodes because the data need to be reordered by time and by baseline and then copied to the global GPU memory before processing. This extra time cost becomes less remarkable on large ($N>=5$) numbers of nodes. We should mention the apparent relationship between the running time and $N_{\rm node}$ could be biased by the limited data and image sizes which are rather small in these benchmark tests. The parallel {\it data loading} at the SKA scale would not depend on the number of nodes after the splitting in time is done.

(5) Data precision exerts a significant influence on MPI+CUDA {\it gridding} processing. Single-precision computation is prominently faster than that for double-precision. The precision has little effect on CPU-based MPI+OpenMP processing. In all these experiments, single-precision GPU-based processing shows the best computational performance with least time consumption.

(6) The {\it gridding} runtime shows a remarkable increase when the support size of the convolution kernel exceeds 8 because more loops are included in the processing. The {\it gridding} runtime continues to steadily increase with image size from 128 pixels to 2048 pixels; after that, the runtime rapidly increases. This should be kept in mind because SKA survey image sizes are usually larger than 8192 pixels. GPU-based MPI+CUDA processing shows the most stable variation.
The experimental results and experience gained from the present work are helpful for the imaging pipeline development of not only the SKA, but also other radio interferometers.

\section{Acknowledgments}\label{sec:6}
This work used China SKA Regional Center prototype system at Shanghai Astronomical Observatory, funded by the National Key R\&D Programme of China (under grant number  2018YFA0404603) and Chinese Academy of Sciences (under grant number 114231KYSB20170003). This research also used resources of National Supercomputer Centre in Guangzhou. This work was supported by resources provided by the Pawsey Supercomputing Centre with funding from the Australian Government and the Government
of Western Australia. This work is partly supported by National Natural Science Foundation of China (No. U1831204, 11703069), the Guangxi Cooperative Innovation Center of cloud computing and Big Data (No. 1716), and the Guangxi Colleges and Universities Key Laboratory of cloud computing and complex systems. BQL thanks Chen Wu for helpful discussion and comments on the manuscript.

All authors contributed to the manuscript preparation. Baoqiang Lao performed the experiment and data analysis, and contributed to the paper writing. Tao An is responsible for the design of the experiment, coordinated the group work and drafted the paper. Ang Yu contributed to the experiment implementation. Wenhui Zhang and Junyi Wang participated in the technical definition of the experiment. Quan Guo, Shaoguang Guo and Xiaocong Wu contributed to the preparation of the experiment.

\medskip

\begin{wrapfigure}{l}{25mm}
    \includegraphics[width=1.25in,height=1.5in,clip,keepaspectratio]{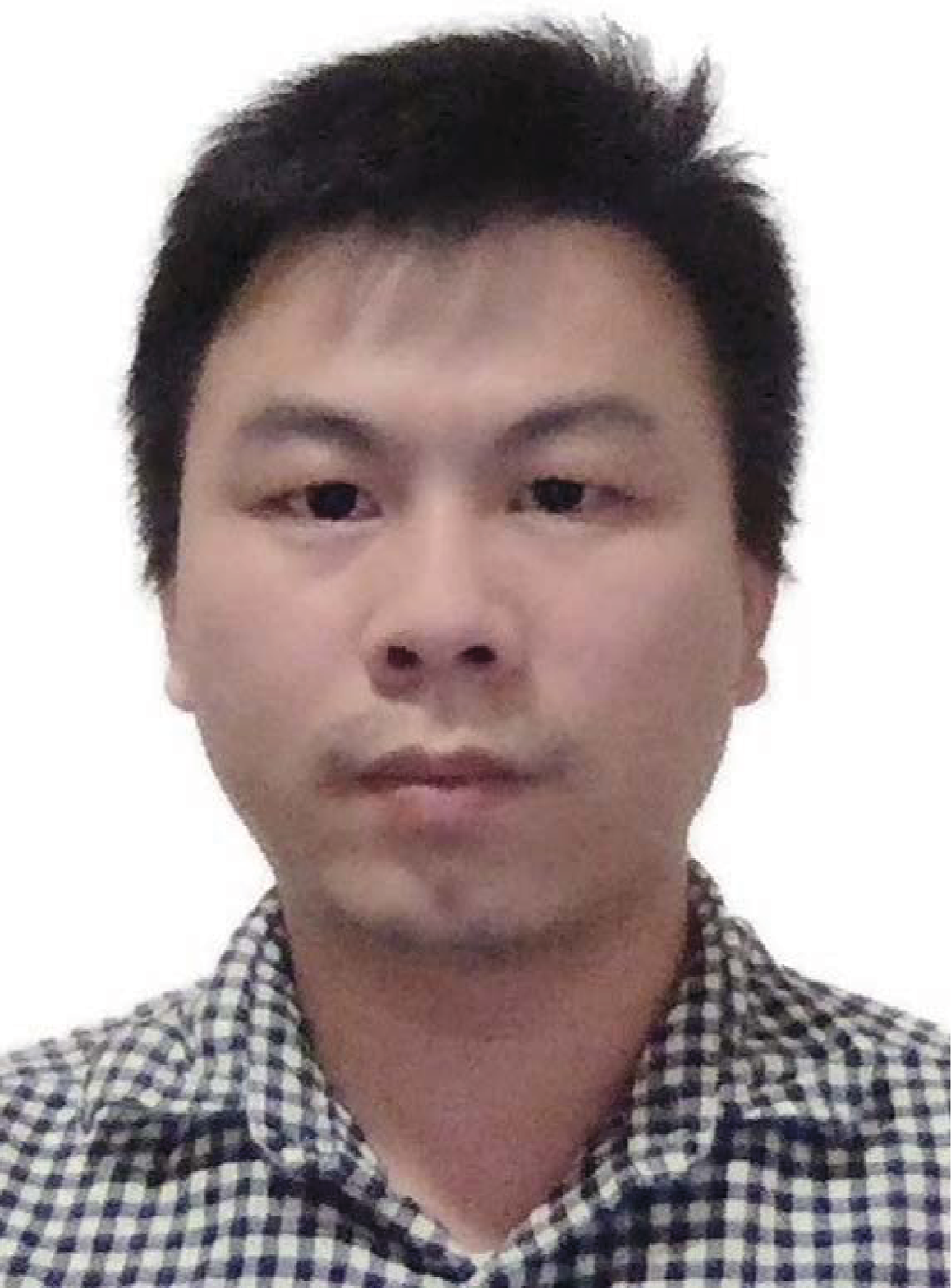}
  \end{wrapfigure}\par
  \textbf{Baoqiang LAO} received the M.Sc. degree from Guilin University of Electronic Technology, Guilin, China, in 2015, in Information and Communication Engineering. He is currently a software engineer of Shanghai Astronomical Observatory (SHAO) of the Chinese Academy of Sciences and a number of SHAO's Square Kilometre Array (SKA) group. His research area mainly focuses on SKA continuum survey imaging, wide-field imaging, HPC performance computing, parallel programing and data intensive astronomy technique.\par

\begin{wrapfigure}{l}{25mm}
    \includegraphics[width=1.25in,height=1.5in,clip,keepaspectratio]{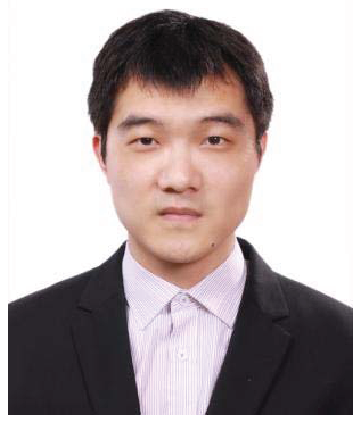}
  \end{wrapfigure}\par
  \textbf{Tao AN}, professor at the Shanghai Astronomical Observatory of the Chinese Academy of Sciences. Research interests: high resolution imaging of compact astronomical objects including black hole, jets, pulsars. He is a member of Square Kilometre Array (SKA) Regional Centre Streering Committe, and responsible for the China SKA Data Center preparation. He also serves in the International Astronomical Society (IAU) Commission B4 (Radio Astronomy), European VLBI Programme Committee.\par

\end{multicols}
\end{document}